\documentclass[journal=tches,preprint]{iacrtrans}

\usepackage{amsmath,amssymb}
\usepackage{booktabs}
\usepackage{multirow}
\usepackage{soul,xcolor}
\usepackage[group-separator={,},group-minimum-digits=4]{siunitx}
\usepackage{tikz}
\usetikzlibrary{arrows.meta,positioning,calc,fit,backgrounds,decorations.pathreplacing,shapes.geometric}
\usepackage{algorithm}
\usepackage{algpseudocode}
\usepackage[dvipsnames]{xcolor}
\usepackage{url}

\newcommand{\lib}{Zeppelin}                  % library name (single point of change)
\newcommand{\cyc}[1]{\num{#1}}               % cycle counts and other grouped numbers
\newcommand{\code}[1]{\texttt{#1}}

\newcommand{\smul}{\cdot}

\newcommand{\barcellL}[3]{% left-aligned text, bar grows from the left
  \begin{tikzpicture}[baseline=-0.5ex]
    \fill[#3] (0,-0.09) rectangle (#2,0.09);
    \node[anchor=west, font=\small] at (0,0) {#1};
  \end{tikzpicture}%
}
\newcommand{\barcellR}[3]{% right-aligned text, bar still grows from the left
  \begin{tikzpicture}[baseline=-0.5ex]
    \fill[#3] (0,-0.09) rectangle (#2,0.09);
    \node[anchor=east, font=\small] at (1.8,0) {#1};
  \end{tikzpicture}%
}

\tikzset{
  zbox/.style={draw, rounded corners=2pt, fill=black!4, align=center,
               inner sep=4pt, font=\small},
  zop/.style={zbox, fill=blue!7},            % computation
  zsec/.style={zbox, fill=orange!14},        % secret-bearing
  zpub/.style={zbox, fill=green!9},          % public data
  zmem/.style={zbox, fill=yellow!14},        % storage
  zarr/.style={-{Latex[length=2mm]}, line width=0.7pt},
  zret/.style={zarr, dashed},
  znote/.style={font=\footnotesize, align=center},
  zhel/.style={zbox, fill=blue!16},    % Helium/MVE four-lane
  zcpu/.style={zbox, fill=black!6},    % scalar Cortex-M55
  zhw/.style={zbox,  fill=orange!18},  % CRYP+DMA peripheral
  zfr/.style={draw, rounded corners=3pt},
}

\title[\lib: Client-Side BFV for Helium-Powered MCUs]{Zeppelin: Client-Side BFV Encryption and Decryption for Helium-Powered Microcontrollers}

\author[Muhammad Saif ul Islam et al.]{Muhammad Saif ul Islam \and Karar Haider \and Manaal Malik \and Naveed Anwar Bhatti \and Muhammad Ali Siddiqi}
\institute{School of Science and Engineering, Lahore University of Management Sciences, Lahore, Pakistan}

\begin{document}

\maketitle

\begin{abstract}
The growth of the Internet of Things (IoT) has increased concerns over the privacy of data collected by resource-constrained sensing devices. Homomorphic encryption (HE) is an attractive way to address this: a device could encrypt its data once and let an untrusted cloud server compute on the ciphertext without seeing the underlying values. In practice, HE's memory and computational cost have kept it out of reach of microcontroller-class devices. Prior work, SEAL-Embedded, made this feasible for the first time using the CKKS scheme, but left three opportunities unexplored: its arithmetic is entirely scalar, even on hardware with a vector instruction set; it never decrypts on the device, so the client can produce ciphertexts but not consume a result; and it does not explore BFV, a scheme whose encoding uses only integer arithmetic and decrypts exactly, which enables applications previously impossible given CKKS's inherent approximation. We present Zeppelin, the first HE library to use an embedded vector instruction set, and the first to support \textit{both} encryption and decryption on an embedded device, using BFV, which had not previously been demonstrated on MCU-class hardware at all. Zeppelin vectorizes the number-theoretic transform, HE's main computational bottleneck, for ARM's Helium extension, and includes a decryption procedure built to avoid large-integer arithmetic and any timing leakage of the secret key. A server-side adapter converts Zeppelin's ciphertexts into a format compatible with Microsoft SEAL, giving an end-to-end deployment where the device handles encryption and decryption while the server performs all homomorphic computation. On an STM32N6 MCU with an ARM Cortex-M55, Zeppelin encodes and encrypts 4096 packed values in 4.76\,ms (128-bit security, per the HE standard) and decrypts and decodes the result in 5.38\,ms, using under 500\,KB of RAM, with the vectorized NTT engine ${\sim}2.7\times$ faster than an equivalent scalar implementation on the same core.
\end{abstract}
% =====================================================================
\section{Introduction}
\label{sec:intro}

Outsourcing computation to the cloud typically means giving up confidentiality: the server must see data in the clear to operate on it. Homomorphic encryption (HE) removes this trade-off by allowing computation directly on encrypted data, so a client can offload work to an untrusted server without ever revealing its inputs. This property is particularly valuable for Internet-of-Things (IoT) deployments, where sensing devices are resource-constrained and routinely offload computation to the cloud, yet increasingly handle sensitive data that the device owner cannot afford to expose. Bringing HE onto such devices, however, requires a scheme whose client-side operations fit within a microcontroller's tight memory and compute budget.

CKKS, an approximate-arithmetic HE scheme with native support for real-valued data, is a natural target for sensor-style workloads, and SEAL-Embedded~\cite{sealembedded} demonstrated that this is feasible: CKKS encoding and encryption fit within 64--137\,KB of RAM, in 737\,ms on a Cortex-M4 and 77\,ms on a Cortex-A7. It leaves two things unaddressed, however. First, its arithmetic is entirely scalar: even on the Cortex-A7, whose core includes the NEON vector extension, SEAL-Embedded computes the number-theoretic transform (NTT) in plain scalar C and so leaves the throughput NEON could offer untapped, since the NTT dominates HE's cost and is exactly the workload an embedded vector extension can accelerate. Second, it never decrypts on-device; the client can produce ciphertexts for the server but cannot itself consume a computed result without an additional, unimplemented decryption step. Consider, for example, a wearable that offloads risk scoring to the cloud: it still needs some other trusted party to decrypt the result before the patient can act.

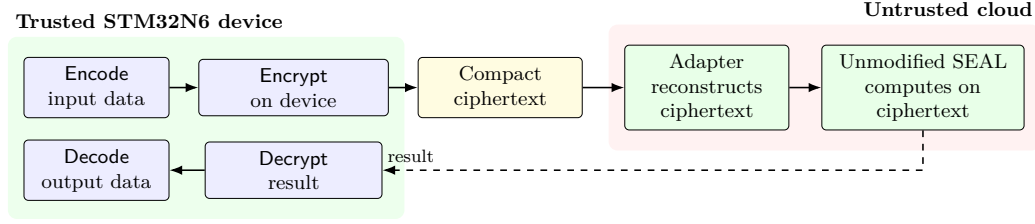
\begin{figure}[t]
\centering
\resizebox{\linewidth}{!}{%
\begin{tikzpicture}[font=\small]
  \node[zop, text width=2cm] (encode) at (0,0.75) {\textsf{Encode}\\input data};
  \node[zop, text width=2.7cm] (encrypt) at (3.1,0.75) {\textsf{Encrypt}\\on device};
  \node[zop, text width=2.5cm] (dec) at (3.1,-0.55) {\textsf{Decrypt}\\result};
  \node[zop, text width=2cm] (decode) at (0,-0.55) {\textsf{Decode}\\output data};
  \node[zmem, text width=2.3cm] (wire) at (6.35,0.75) {Compact\\ciphertext};
  \node[zpub, text width=2.3cm] (adapter) at (9.6,0.75) {Adapter\\reconstructs\\ciphertext};
  \node[zpub, text width=2.9cm] (seal) at (13.0,0.75) {Unmodified SEAL\\computes on\\ciphertext};
  \begin{scope}[on background layer]
    \node[fill=green!7, rounded corners, fit=(encode)(encrypt)(dec),
          inner xsep=7pt, inner ysep=9pt] (dev) {};
    \node[fill=red!5, rounded corners, fit=(adapter)(seal),
          inner xsep=7pt, inner ysep=9pt] (cloud) {};
  \end{scope}
  \node[anchor=south west, font=\bfseries\small] at (dev.north west) {Trusted STM32N6 device};
  \node[anchor=south east, font=\bfseries\small] at (cloud.north east) {Untrusted cloud};
  \draw[zarr] (encode) -- (encrypt);
  \draw[zarr] (encrypt) -- (wire);
  \draw[zarr] (wire) -- (adapter);
  \draw[zarr] (adapter) -- (seal);
  \draw[zret] (seal.south) |- ($(dec.east)+(0.9,0)$) -- node[above, font=\footnotesize]{result} (dec.east);
  \draw[zarr] (dec) -- (decode);
\end{tikzpicture}%
}
\vspace{-0.6cm}
\caption{\lib{} end-to-end deployment.}
\label{fig:endtoend}
\end{figure}

A third gap sits alongside the two above, in the choice of scheme itself. CKKS encodes through a double-precision complex FFT, which is slow and memory-heavy on a fixed-point MCU. BFV~\cite{bra12,fv12} encodes with an integer NTT modulo the plaintext modulus $t$, reusing the same modular arithmetic units that encryption already requires, and packs twice as many slots at equal ring degree while returning exact integers rather than approximations. Despite this apparent fit for MCU-class hardware, no BFV encode-and-encrypt library has been demonstrated on such a device.

Zeppelin\footnote{Zeppelin is available at: \url{https://anonymous.4open.science/r/zeppelin-2F7C}} closes all three gaps at once: it uses a vector ISA to accelerate the NTT, and it runs the entire BFV client pipeline, encoding, key generation, encryption, and decryption, on the device itself, with evaluation remaining server-side as is standard for leveled HE deployments.
In essence, this paper makes the following contributions:
\begin{itemize}
    \item \textbf{The first HE implementation on an embedded vector ISA.} Since the NTT dominates HE's computational cost, we design a blocked NTT schedule for ARM's Helium/MVE extension that factors a 4096-point transform into tiles that fit the vector register file and instruction memory, together with a lazy-reduction scheme that avoids unnecessary modular reductions between stages.

    \item \textbf{The first on-device decryption for embedded HE, and the first BFV client.} Zeppelin's on-device decryption closes the loop that a complete client-side HE implementation requires. The library also provides on-device encoding, key generation, and symmetric and public-key encryption, made practical by targeting a vector MCU, using a division-free decryption procedure and a compact seed-based ciphertext format.

    \item \textbf{An end-to-end deployment compatible with existing HE tooling.} A server-side adapter converts the ciphertext bytes Zeppelin transmits into ciphertexts compatible with the unmodified Microsoft SEAL library~\cite{seal}, allowing an untrusted server to perform homomorphic computation (Figure~\ref{fig:endtoend}).

    \item \textbf{A rigorous, cross-implementation correctness methodology.} We check a reference implementation, an instruction-accurate simulator, and real hardware against each other, and confirm interoperability by showing that unmodified Microsoft SEAL can decrypt our library's ciphertexts and correctly evaluate homomorphic additions, multiplications, and relinearizations over them.

    \item \textbf{Measured performance on a real embedded device.} We provide performance, memory usage, and an approximate energy estimate for our library on an ARM Cortex-M55 microcontroller, at n = 4096 with a 104-bit ciphertext modulus (128-bit security).
    
\end{itemize}

The rest of the paper proceeds as follows. Section~\ref{sec:prelim} covers background and related work. Section~\ref{sec:design} describes Zeppelin's design and threat model. Section~\ref{sec:impl} walks through the implementation as design decisions, each with rationale and measured impact. Section~\ref{sec:eval} presents the evaluation, and Section~\ref{sec:concl} concludes the discussion.

% =====================================================================
\section{Preliminaries}
\label{sec:prelim}

\subsection{Notation} We write $n$ for the ring dimension, a power of two, and $R = \mathbb{Z}[x]/(x^n+1)$ for the corresponding $2n$-th cyclotomic ring, the standard setting for negacyclic polynomial arithmetic and NTT-based multiplication. 
For a modulus $q$, $R_q = R/qR$ denotes $R$ with every coefficient additionally reduced modulo $q$ and represented as an unsigned integer in $[0, q-1]$; this is the ring used throughout the paper for ciphertexts, keys, and intermediate values, since bounding coefficients to a fixed modulus is what keeps them representable in machine words and gives the ring the finite structure RLWE's security and the NTT's fast multiplication both rely on. We write $[\cdot]_q$ for this reduction. This notation applies to any modulus in place of $q$, for instance $R_t$ or $R_Q$ once the plaintext modulus $t$ and ciphertext modulus $Q$ are introduced in Section~\ref{sec:bfv}.

Three distributions recur throughout the paper, each extended to polynomials by applying it independently to every one of the $n$ coefficients. $\mathcal{U}(S)$ denotes the uniform distribution over a set $S$; in particular $\mathcal{U}(R_q)$ samples a polynomial by drawing each coefficient independently and uniformly from $[0, q-1]$. We write $\chi_{\mathrm{ter}}$ for the ternary distribution that draws each coefficient independently and uniformly from $\{-1, 0, 1\}$, and $\mathcal{D}_\sigma$ for the discrete Gaussian distribution over $\mathbb{Z}$ with standard deviation $\sigma$, applied coefficient-wise to sample a polynomial's error terms. Throughout the paper $n = 4096$ and the plaintext modulus is $t = 65537$; $\mathrm{br}_6(\cdot)$ denotes 6-bit bit-reversal (Section~\ref{sec:seeded}).

\subsection{Leveled Homomorphic Encryption and RLWE}
Homomorphic encryption (HE) allows a party to compute directly on encrypted data, without ever decrypting it or exposing it to the party performing the computation. This paper works with \emph{leveled} HE: ciphertexts support both addition and multiplication, but only up to a multiplicative depth fixed in advance by the choice of parameters, rather than schemes that use bootstrapping to refresh noise and support unbounded computation.
An HE scheme is typically organised around six operations: parameter selection, key generation, encoding, encryption, evaluation, and decryption. \textbf{Parameters} fix the ring dimension and modulus needed for a target security level and multiplicative depth; \textbf{key generation}, run by the trusted party holding the secret key, derives that secret key together with any public encryption keys and evaluation keys needed downstream; \textbf{encoding} converts a vector of input values into the plaintext polynomial the scheme operates on; \textbf{encryption}, which may be symmetric (using the secret key) or asymmetric (using a public key), is randomised so that encrypting the same message twice yields different ciphertexts; \textbf{evaluation} applies homomorphic operations to ciphertexts, typically carried out by an untrusted party that never sees the underlying plaintext; and \textbf{decryption}, again restricted to the trusted party, recovers the plaintext from a ciphertext. Zeppelin instantiates encoding, key generation, encryption, and decryption on the device, while evaluation runs on the untrusted server, as described in Section~\ref{sec:design}.

The efficient HE schemes in widespread use today, such as BFV~\cite{bra12,fv12} and CKKS~\cite{ckks17}, all derive their security from the Ring Learning with Errors (RLWE) problem~\cite{lpr10}.
For a secret $s \leftarrow \chi_{\mathrm{ter}}$, an RLWE instance samples a uniformly random $a \leftarrow \mathcal{U}(R_q)$ and a small error $e \leftarrow \mathcal{D}_\sigma$, then computes $b = [as+e]_q$; the RLWE assumption states that the resulting pair $(b, a)$ is computationally indistinguishable from a pair drawn uniformly at random from $R_q^2$. We adopt the community-standard choices for both randomised quantities, $\chi_{\mathrm{ter}}$ for the secret key and $\mathcal{D}_\sigma$ with $\sigma = 3.2$ for the error, and select parameters within the classical security envelope of the HE security standard~\cite{hestd18}.

\subsection{BFV}
\label{sec:bfv}
BFV was introduced by Fan and Vercauteren~\cite{fv12} as a scale-invariant variant of Brakerski's earlier scheme~\cite{bra12}, and unlike CKKS, it is an exact scheme: every homomorphic operation on ciphertexts corresponds exactly to the same operation on the underlying integers modulo $t$, with no approximation error introduced by the scheme itself. Where CKKS truncates low-order bits of a floating-point value to keep ciphertexts within a bounded dynamic range, BFV instead requires that computed values never exceed the plaintext modulus $t$; if they do, the result wraps around modulo $t$ rather than degrading gracefully. This exactness makes BFV the natural choice for computations where an approximate answer is not acceptable, private set intersection, exact counting and aggregation, and encrypted database queries among them, and is the reason BFV, alongside BGV, remains the standard choice in application settings where CKKS's approximation is unacceptable.

This exactness comes with a practical consequence for sensor-style IoT data, which is naturally real-valued: since BFV's plaintext space is $\mathbb{Z}_t^n$, not $\mathbb{R}^{n/2}$, real-valued readings must first be quantized to fixed-point integers, by scaling and rounding, before encoding, rather than encoded directly the way CKKS allows. We choose BFV despite this, since the two schemes trade one cost for another: CKKS's encoder needs a double-precision complex FFT, expensive and memory-heavy on a fixed-point MCU without a hardware floating-point pipeline sized for it, while BFV's encoder is a modular integer NTT, reusing the same modular arithmetic the library already needs for ciphertext operations.

Since $t \equiv 1 \pmod{2n}$, $\mathbb{Z}_t$ contains a primitive $2n$-th root of unity $\omega_t$, and the map $\sigma_t : R_t \to \mathbb{Z}_t^n$ given by $\sigma_t(m) = (m(\omega_t^{2i+1}))_{i=0}^{n-1}$ is a ring isomorphism, the integer-arithmetic analogue of CKKS's canonical embedding. 
This isomorphism packs $n$ independent integer slots into a single plaintext polynomial (called \textit{batching}), twice as many as CKKS's $n/2$ slots at equal ring degree, so that a single homomorphic operation acts on all $n$ values at once; batching is computed in practice as an NTT modulo $t$~\cite{sv14}.

\textbf{BFV.Encode}($z \in \mathbb{Z}_t^n$): output the plaintext polynomial $m = \sigma_t^{-1}(z) \in R_t$, computed as an inverse NTT modulo $t$.

\textbf{BFV.KeyGen}: sample the secret key $s \leftarrow \chi_{\mathrm{ter}}$. For the public key, sample $a \leftarrow \mathcal{U}(R_Q)$ and $e \leftarrow \mathcal{D}_\sigma$, and output $\mathrm{pk} = (p_0, p_1) = ([-(as+e)]_Q,\ a) \in R_Q^2$.

\textbf{BFV.SymEncrypt}(secret key $s \leftarrow \chi_{\mathrm{ter}}$, plaintext $m \in R_t$): sample $a \leftarrow \mathcal{U}(R_q)$ and $e \leftarrow \mathcal{D}_\sigma$. Output $(c_0, c_1) = ([\Delta m - as - e]_Q,\ a) \in R_Q^2$, where $\Delta = \lfloor Q/t \rfloor$.

\textbf{BFV.AsymEncrypt}(public key $\mathrm{pk} = (p_0, p_1) \in R_Q^2$, plaintext $m \in R_t$): sample $u \leftarrow \chi_{\mathrm{ter}}$ and $e_1, e_2 \leftarrow \mathcal{D}_\sigma$. Output $(c_0, c_1) = ([p_0 u + e_1 + \Delta m]_Q,\ [p_1 u + e_2]_Q)$.

Unlike SEAL-Embedded, which omits key generation, evaluation, and decryption from its scheme description since these are not implemented on its target device, our library also performs key generation and decryption on-device, with only evaluation remaining server-side. On-device decryption is one of Zeppelin's central contributions; we defer its description to Section~\ref{sec:decrypt}, where the division-free BEHZ correction needed to implement it efficiently in RNS is presented alongside its rationale.

\subsection{Residue Number System}
\label{sec:rns}
Representing a 104-bit modulus $Q$ directly would require multi-precision arithmetic our 32-bit core does not have. We avoid this with a Residue Number System (RNS): by the Chinese Remainder Theorem, $R_Q \cong R_{q_0} \times \cdots \times R_{q_{L-1}}$, so a polynomial modulo $Q$ can be represented instead as $L$ independent polynomials, one modulo each prime $q_k$ (here $L = 4$). Encoding, sampling, NTTs, pointwise multiplication, and ciphertext assembly all operate on these residues independently, as $L$ parallel per-prime loops, with no communication between primes. The one exception is decryption's scale-and-round step, which must combine information across all $L$ residues to recover a single result modulo $t$. 
For this step we use the integer-only BEHZ base-conversion correction~\cite{behz16}, the same variant Microsoft SEAL implements, rather than the floating-point correction of the HPS variant~\cite{HPS19}, keeping decryption consistent with the integer-only arithmetic used throughout the rest of the library. We describe our specialisation of this correction in Section~\ref{sec:decrypt}.

\subsection{Number Theoretic Transform} 
\label{sec:nttprelim}
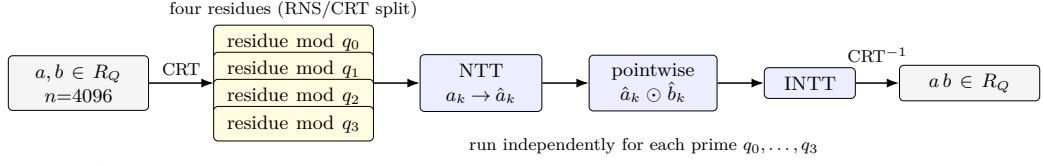
\begin{figure}[t]
\centering
\resizebox{\linewidth}{!}{%
\begin{tikzpicture}[font=\small]
  \node[zbox, text width=2.0cm] (poly) at (0,0) {$a,b\in R_Q$\\$n{=}4096$};
  \foreach \k/\y in {0/0.66,1/0.22,2/-0.22,3/-0.66}
    \node[zmem, minimum width=2.6cm, minimum height=0.36cm] (r\k) at (3.5,\y)
      {residue $\bmod\ q_{\k}$};
  \node[znote, anchor=south] at (3.5,0.92) {four residues (RNS/CRT split)};
  \node[zop, text width=1.7cm] (ntt) at (6.55,0) {NTT\\$a_k\!\to\!\hat a_k$};
  \node[zop, text width=1.8cm] (pw) at (9.35,0) {pointwise\\$\hat a_k\odot\hat b_k$};
  \node[zop, text width=1.2cm] (intt) at (11.9,0) {INTT};
  \node[zbox, text width=2.0cm] (out) at (14.5,0) {$a\,b\in R_Q$};
  \draw[zarr] (poly.east) -- node[above, font=\footnotesize]{CRT} (2.20,0);
  \draw[zarr] (4.80,0) -- (ntt.west);
  \draw[zarr] (ntt) -- (pw);
  \draw[zarr] (pw) -- (intt);
  \draw[zarr] (intt.east) -- node[above, yshift=6pt, font=\footnotesize]{CRT$^{-1}$} (out.west);
  \node[znote] at (9.2,-1.02)
    {run independently for each prime $q_0,\dots,q_3$};
\end{tikzpicture}%
}
\vspace{-.7cm}
\caption{The generic mechanism \lib{} uses for \textbf{polynomial multiplication}, for computing $a \cdot s$ during encryption and $c_1 \cdot s$ during decryption. A polynomial in $R_Q$ is represented as four independent residues (Section~\ref{sec:rns}), and multiplying two such polynomials reduces, within each residue independently, to a forward NTT, a pointwise multiply, and an inverse NTT. \lib{} vectorises this per-prime pipeline.}
\label{fig:prelim}
\end{figure}

The dominant cost in HE, and in Zeppelin, is negacyclic polynomial multiplication: transforming two operands into the NTT domain, multiplying them pointwise, and transforming the result back, an $O(n \log n)$ procedure built from per-coefficient modular products. As is standard practice, twiddle factors are precomputed in Montgomery form~\cite{montgomery85,ln16}, avoiding repeated modular reductions during the transform itself. Our forward kernels take natural-order input and emit bit-reversed output in the Montgomery domain (a decimation-in-frequency, or DIF, structure); inverse kernels reverse this, taking bit-reversed input and emitting natural-order output in $[0, q_k)$ (decimation-in-time, or DIT). Algorithm~\ref{alg:ntt} gives the reference transform this hardware implementation computes. Figure~\ref{fig:prelim} shows how this transform fits into the per-prime pipeline Zeppelin vectorises.

\subsection{Related Work}
\label{sec:related}
\textbf{Embedded HE.} SEAL-Embedded~\cite{sealembedded} is, to date, the only prior HE library targeting MCU-class devices: it encodes and encrypts CKKS on Cortex-M4 and Cortex-A7 cores, using scalar arithmetic and an aggressive memory reuse scheme, with decryption performed server-side only. Zeppelin differs along several axes: scheme (BFV's integer encoder rather than CKKS's floating-point FFT), arithmetic (a Helium/MVE vector NTT rather than scalar code), and scope (on-device decryption, and an NTT-domain sampling optimisation built on SEAL's existing seeded ciphertext format). SEAL-Embedded is optimised for memory on the smallest cores, while Zeppelin targets the emerging class of Helium-equipped devices, deliberately spending memory to reduce latency, a trade-off quantified in Section~\ref{sec:eval-mem}.
Because device class, memory budget, scheme, and ISA all differ from SEAL-Embedded at once, we do not treat that comparison as evidence for vectorisation specifically; Section~\ref{sec:eval-ablation} isolates the vectorised engine's contribution with a same-core, same-image ablation instead.
A separate family of schemes, TFHE and its relatives~\cite{tfhe20}, targets Boolean and gate-level workloads under per-gate bootstrapping; this is a different computational regime from the packed, leveled arithmetic Zeppelin and SEAL-Embedded both target, and we do not compare against it directly.

\textbf{Vector NTTs for post-quantum cryptography.} Embedded vector NTTs are well developed for lattice-based post-quantum cryptography (PQC) at the much smaller ring dimension $n=256$: Helium/MVE implementations for Kyber and Dilithium culminate in the SLOTHY superoptimiser and its pqmx kernels~\cite{pqmx}, with a corresponding Neon line for A-profile cores~\cite{neonntt22} and scalar Cortex-M4 baselines in pqm4~\cite{pqm4,kyber18,dilithium18}. Vectorised NTTs for HE itself exist as well, but off the embedded platform: Rodrigues et al.~\cite{rvvntt} accelerate HE's NTT using the RISC-V Vector extension on server-class processors, not an MCU-class embedded vector ISA.
To the best of our knowledge, Zeppelin is the first HE implementation to target an embedded vector ISA. HE poses a different engineering problem than PQC at a different scale: at $n=256$, the PQC kernels merge an entire transform within the Helium register file, whereas HE's $n=4096$ residues are far larger than that register file or the available instruction memory can hold at once, forcing a blocked, multi-pass schedule rather than a single merged kernel.

% =====================================================================
\section{\lib{} Design}
\label{sec:design}

\lib{} has two components: the \emph{device library} (encode, key generation, symmetric and public-key encryption, decryption) and a \emph{cloud adapter} that converts the device's residue-major serialized bytes into ciphertexts of an unmodified Microsoft SEAL~\cite{seal} instance for server-side evaluation; the adapter itself holds no secrets (see Figure~\ref{fig:endtoend}).

\subsection{Threat Model and Security Preconditions}
\label{sec:threat}

We consider the device, the trusted embedded client that collects and encrypts data, and the server, an untrusted party to which the device offloads computation, as the two parties in our system; the device's owner controls it and holds its secret key, while the server is assumed to follow the protocol correctly but to try to learn as much as it can from what it observes, a standard \emph{semi-honest} assumption. HE is what lets this arrangement work at all: it protects the confidentiality of the device's data against the server, since the server never has access to anything but ciphertexts and never needs the secret key to operate on them. Transport integrity, keeping traffic from being tampered with in transit, is assumed to come from the usual session-layer protections (e.g.\ TLS) rather than from HE itself, which is designed to protect confidentiality, not to replace a secure channel.

Since the device generates its own secret and public keys locally via key generation, no external key-provisioning channel is needed; we do assume the device's software and physical integrity are not compromised, since an attacker with such access could already reach the underlying sensor data directly, and we treat device compromise, including extraction of the on-device secret key, as out of scope.
\lib{}'s on-device \code{KeyGen} produces only the secret and public keys; it neither generates nor uses relinearization or Galois/rotation keys. Relinearization is instead demonstrated (Section~\ref{sec:correctness}) using SEAL's own key generator, run once at deployment time, over the imported secret key, on the trusted side (e.g.\ a separate trusted server); only the resulting, safe-to-share keys reach the untrusted server, and we do not report their (off-device) generation cost.

On-device, we defend the \emph{timing} side channel specifically: no secret-dependent branch, no secret-indexed memory access, and no division in secret-handling code, a property we argue at the source and ISA level and verify on emitted assembly (Section~\ref{sec:ct}). Power and electromagnetic side channels, and fault attacks, are out of scope; mitigating them, whether through software masking or hardware countermeasures, is left as future work.

The seeded ciphertext format (Section~\ref{sec:seeded}) introduces one additional precondition, worth stating here as part of the design rather than as an implementation detail. Classic BFV already transmits the uniform mask $a$ verbatim as part of the ciphertext, so $a$ is public either way, publishing a seed instead, one that expands deterministically through a public PRF (AES-128-CTR) to the same distribution, reveals nothing an attacker did not already have: RLWE's hardness never rested on $a$ being secret, only on it being uniformly random, a property the unchanged, bias-free rejection sampler still preserves when $a$ is derived from a public seed. What \emph{must} hold instead is that the error keystream is never derivable from that published seed; if it were, an attacker could recover $s$ directly from $c_0 = \Delta m - as - e$. 
\lib{} enforces this separation structurally rather than by convention: seeded encryption draws from two independent PRNG states, \code{prng\_a}, whose seed is fresh from the TRNG for every ciphertext and is the one that gets transmitted, and \code{prng\_e}, whose seed is likewise fresh per ciphertext but never leaves the device.

\subsection{Library Overview}
\label{sec:overview}
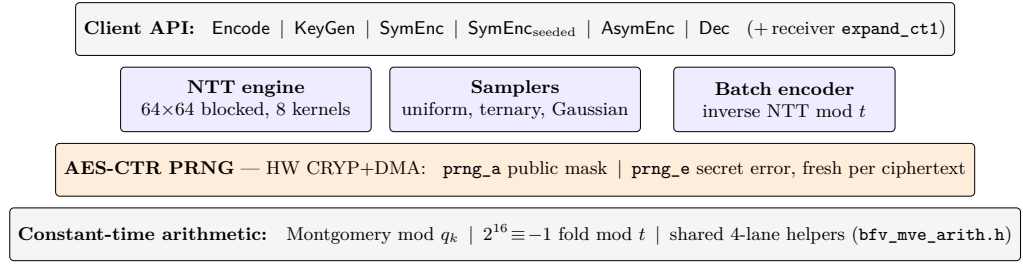
\begin{figure}[t]
\centering
\resizebox{0.98\linewidth}{!}{%
\begin{tikzpicture}[font=\small]
  \node[zbox, minimum width=13cm, minimum height=0.9cm] (top) at (7,4.7)
    {\textbf{Client API:}\;\; \textsf{Encode} \,$\vert$\, \textsf{KeyGen}
     \,$\vert$\, \textsf{SymEnc} \,$\vert$\, \textsf{SymEnc}$_{\mathrm{seeded}}$
     \,$\vert$\, \textsf{AsymEnc} \,$\vert$\, \textsf{Dec}
     \;\;(+\,receiver \code{expand\_ct1})};
  \node[zop, text width=4cm, minimum height=1.1cm] (ntt) at (2.4,3.5)
    {\textbf{NTT engine}\\$64{\times}64$ blocked, 8 kernels};
  \node[zop, text width=4cm, minimum height=1.1cm] (samp) at (7,3.5)
    {\textbf{Samplers}\\uniform, ternary, Gaussian};
  \node[zop, text width=3.5cm, minimum height=1.1cm] (enc) at (11.6,3.5)
    {\textbf{Batch encoder}\\inverse NTT mod $t$};
  \node[zsec, minimum width=13cm, minimum height=0.9cm] (prng) at (7,2.3)
    {\textbf{AES-CTR PRNG} --- HW CRYP$+$DMA:\;\;
     \code{prng\_a} public mask \,$\vert$\, \code{prng\_e} secret error,
     fresh per ciphertext};
  \node[zbox, minimum width=13cm, minimum height=0.9cm] (base) at (7,1.2)
    {\textbf{Constant-time arithmetic:}\;\; Montgomery mod $q_k$ \,$\vert$\,
     $2^{16}\!\equiv\!-1$ fold mod $t$ \,$\vert$\,
     shared 4-lane helpers (\code{bfv\_mve\_arith.h})};
\end{tikzpicture}%
}
\vspace{-0.1cm}
\caption{\lib{} architecture. Hand-written Helium/MVE assembly is confined to the eight NTT kernels; everything else is C over one small set of proven constant-time lane helpers. \code{expand\_ct1}, run by the receiver, reconstructs the second ciphertext half from its transmitted seed. The AES-CTR PRNG feeds the samplers from two structurally separate states, \code{prng\_a} (public mask) and \code{prng\_e} (secret error, Section~\ref{sec:threat}).}
\label{fig:arch}
\end{figure}

The library is layered (Figure~\ref{fig:arch}), and these layers map one-to-one onto the components the rest of this paper optimises. At the base sits constant-time \textbf{Montgomery arithmetic} modulo each prime $q_k$, together with a $2^{16}\!\equiv\!-1$ limb fold modulo $t$. Directly above it is one small, proven set of \textbf{four-lane Helium helpers} (Section~\ref{sec:arith}), the vector primitives every other module in the library is built from. Three compute engines are built on top of these primitives: the \textbf{vector NTT engine}, a blocked $64\times64$ scheduler over eight per-prime lazy MVE kernels; the three \textbf{samplers}, uniform-by-rejection, ternary, and a constant-time Gaussian CDT; and the integer \textbf{batch encoder}, an NTT computed modulo $t$. An AES-CTR PRNG, split into two structurally separate states, feeds the samplers with keystream. The top layer composes all of this into the library's client-facing operations, including the seeded encrypt and key generation fast path and the receiver-side \code{expand\_ct1}. Parameters are fixed as compile-time presets, and all hand-written assembly is confined to the eight NTT kernels described above.

\subsection{Logical Flow}
\label{sec:flow}

\begin{figure}
    \centering
    \includegraphics[width=1\linewidth]{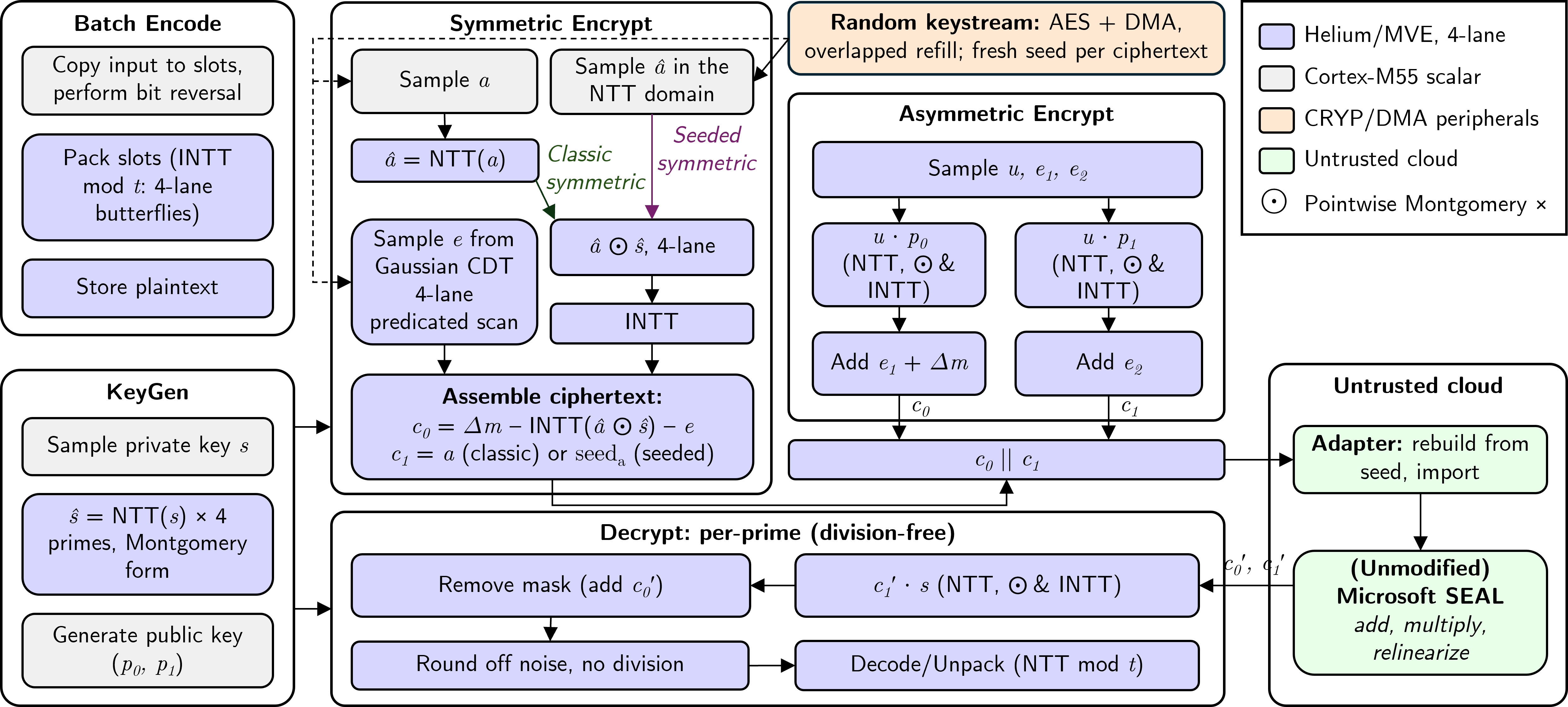}
    \vspace{-0.5cm}
    \caption{\lib{} end-to-end dataflow, coloured by whether each sub-block runs on Helium/MVE or scalar Cortex-M55: encoding, key generation, all three encryption paths, server-side evaluation, and decryption.}
    \label{fig:flow}
\end{figure}

Zeppelin's encryption follows the standard RLWE construction, encrypting zero and adding the encoded plaintext on top, computed independently within each of the four residues (Section~\ref{sec:rns}). The library supports three paths through this construction, classic symmetric encryption, its seeded counterpart, and public-key encryption, all built from the same per-prime engine; Figure~\ref{fig:flow} traces the full dataflow through that engine, encoding, key generation, all three encryption paths, server-side evaluation, and decryption, with every sub-block coloured by the unit that executes it, and marks where the classic path diverges from the seeded one. Of these three, the seeded path is the fastest: it samples the mask $a$ directly in the NTT domain, as $\hat a$, rather than sampling it in the ordinary coefficient domain and then transforming it (Section~\ref{sec:seeded}), so it alone needs no forward NTT on $a$; the inverse NTT that follows the pointwise multiply is still required on every path.

% =====================================================================
\section{Implementation and Optimisations}
\label{sec:impl}

This section walks through \lib{}'s implementation as a series of design decisions, presenting, for each one, what we chose, why, and what effect it had when measured on real hardware. The library was built in two stages: we first got a correctness-first version working, matching a set of known-answer tests bit-for-bit (Section~\ref{sec:correctness} describes this methodology), and only then optimised it, requiring every subsequent change to continue passing those same tests. To make sure each optimisation's reported effect is actually attributable to that change and not to noise, we checked two things at every step: that operations the change should not have touched stayed put, and that any component already optimised and left untouched still reproduced its earlier per-primitive cycle counts (within $0.2\%$). Section~\ref{sec:eval} discusses the resulting sequence of changes and their effects.

\subsection{Low-Level Arithmetic and the Shared Vector Helpers}
\label{sec:arith}

All per-prime coefficient products use Montgomery multiplication~\cite{montgomery85}, with radix $\rho = 2^{32}$ chosen to match the core's native word size; twiddle factors, the per-prime scaling constants $\Delta_k=[\lfloor Q/t\rfloor]_{q_k}$, and every NTT-domain operand are stored in this Montgomery form rather than as ordinary residues. Every reduction that touches secret data ends in a branch-free mask rather than an \code{if}: the final conditional subtraction is computed as \code{t\,+=\,q\,\&\,(t>>63)}, which the compiler emits as straight-line \code{subs;sbc;ands;adds} rather than a branch, a property we verify on the \code{-S} output at every reduction site (Section~\ref{sec:ct} describes a leak this discipline fixed). Reductions modulo $t=2^{16}{+}1$ similarly avoid branching, using an alternating $16$-bit limb fold with masked subtracts; no division instruction appears anywhere on a path that touches secret data.

The C-level vector code is built from one small, proven set of Helium intrinsics helpers, gathered in \code{bfv\_mve\_arith.h}: a four-lane Montgomery multiply (\code{vmulq}/\code{vmulhq} for the low and high products, with the carry applied through lane predication rather than a branch), four-lane canonicalisation and $[0,3q)$ reduction expressed as arithmetic masks (\code{vshrq\_n\_s32} combined with \code{vandq}), and a four-lane mod-$t$ limb fold over $(\mathrm{lo},\mathrm{hi})$ accumulator pairs. Montgomery-form conversion reuses this same multiply primitive, computed as $\mathrm{to\_mont}(x)\equiv\mathrm{mont\_mul}(x,\rho^2\bmod q_k)$. Keeping all of this in one small shared header is deliberate: the further vectorisations of Section~\ref{sec:tier2} are built entirely as compositions of these already-verified lane primitives, so each new port introduces no new constant-time surface to re-verify (Section~\ref{sec:ct}).

\subsection{Random Sampling and PRNG--Compute Overlap}
\label{sec:sampling}

Randomness comes from AES-128-CTR. The uniform polynomial $a$ is drawn by rejection directly in $[0,q_k)$, so its draw count depends only on public data, never on secret values. The ternary secret key $s$ and the ephemeral secret $u$ consume a fixed $4n$ keystream bytes, mapping each coefficient through two fixed-threshold, branch-free comparisons. The Gaussian error $e$ likewise uses a constant-time CDT scan ($\sigma=3.2$, table truncated at $|e|\le21=6.56\sigma$); the fixed-point table quantises each cumulative probability to $2^{-31}$, giving a statistical distance $\lesssim2^{-30}$ per coefficient from the ideal discrete Gaussian, negligible against our security level by the standard analyses of CDT precision~\cite{hprr20}. In the final configuration, this CDT scan is itself four-lane, a predicated Helium scan consuming the identical keystream in the identical order (Section~\ref{sec:tier2}).

Sampling this much randomness is not free, and overlapping it with compute is where our main optimisation here comes from. 
Our profiled baseline spent ${\sim}24$ cycles per keystream byte in the hardware AES peripheral (CRYP) under polling: the core sat waiting on bus traffic to and from the peripheral, with its vector units left completely idle.
We restructured the PRNG around two $\num{20480}$-byte keystream slabs, refilled by the CRYP peripheral through a general-purpose DMA channel while the core continues computing. The slabs sit on exclusive, $32$-byte-aligned cache lines at the front of the PRNG state, with an unconditional invalidate-before-fill and invalidate-after-fill protocol so that DMA and the D-cache can never disagree about a shared line; ahead of every error-sampling window, prefetch hooks trigger the next slab fill early, so a refill is already underway by the time it is needed.
One bring-up fact on the STM32N6~\cite{stm32n6rm} is worth recording here: its resource-isolation framework silently filters non-secure DMA requests for a secure application, so the channel must itself be configured secure-and-privileged.

This restructuring measurably helps most operations (Section~\ref{sec:eval} tabulates the process), though not all: key generation's ternary sampling offers no overlap window, so it pays the slower, synchronous slab rate ($28$ vs.\ $24$ cycles/byte) and slows down, a genuine trade-off rather than measurement noise. PRNG-bearing operations also show a bimodal timing distribution, since any single iteration may or may not include a slab refill; individual measurements cluster around two values accordingly, and we report medians of $1000$ runs throughout to smooth this variation out. No new secret-dependent branch is introduced: the only new control flow branches on a scheduling flag derived from slab geometry and call order, never from key, error, or keystream values (Section~\ref{sec:ct}).

\subsection{Memory Management and Placement}
\label{sec:memory}

\lib{} partitions all polynomial state by RNS prime: encoding, sampling, the NTT, and ciphertext assembly all operate on one $\num{16384}$-byte residue row at a time. Unlike SEAL-Embedded, we do not compress the working set toward a $256$-KB budget: the STM32N6 carries $1536$~KB of AXISRAM, and \lib{} spends this capacity on full per-prime polynomials to keep the vector pipeline fed. We measured the runtime footprint directly, rather than estimating it, using an instrumented build that snapshots every buffer at every processing stage, on the FVP (ARM's instruction-accurate simulator) and again on silicon, with the known-answer tests (KATs, Section~\ref{sec:correctness}) reproduced in the same run so the profile reflects the library's real processing stages rather than a stripped-down test harness. The peak concurrent live set stays well within a third of the available AXISRAM across every operation (Table~\ref{tab:mem}); the library allocates nothing dynamically, and the measured stack high-water mark, worst case the MVE decrypt tail, comfortably fits a $2$~KB reserve. Working-set residency is therefore not the constraint here; the real design problem is \emph{cache} behaviour, and the two placement experiments below, testing whether the NTT's twiddle tables and kernel code should live in the core's small, fast Tightly-Coupled Memory blocks (DTCM for data, ITCM for instructions) rather than ordinary cached memory, were each resolved by measurement, one confirming the ``obvious'' choice, one rejecting it.

\paragraph{Twiddle tables to DTCM: taken.} The $4096$-point transform of Section~\ref{sec:ntt} consumes $16$~KB of twiddle constants per prime and direction, $256$~KB in total with the vector-paired layout. The eight \emph{forward} planes alone come to exactly $128$~KB; at boot, we copy them into DTCM (a NOLOAD section, ECC-scrubbed then byte-verified), leaving the inverse planes in flash-resident \code{.rodata} behind the data cache. Measured directly after this change was adopted, in the same final image, the forward NTT's twiddle pass fell $35\%$, taking the whole forward transform from \cyc{406923} to \cyc{380768} cycles per prime ($-6.4\%$); the gains propagate downstream, with keygen (which runs four forward NTTs) improving $4.4\%$, public-key keygen $2.7\%$, decrypt $1.3\%$, and symmetric encrypt $1.0\%$.

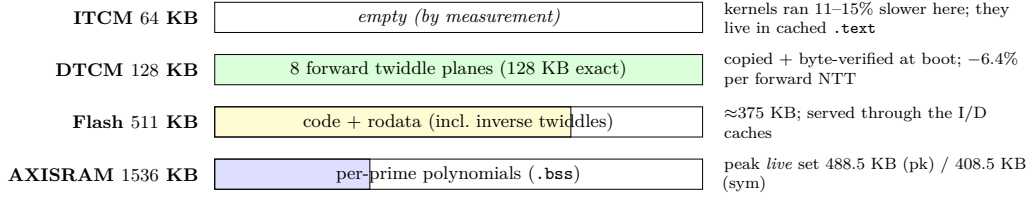
\begin{figure}[t]
\centering
\resizebox{\linewidth}{!}{%
\begin{tikzpicture}[font=\small]
  \def\BW{8.4}
  % ---- ITCM: empty ----
  \node[anchor=east, font=\small\bfseries] at (-0.15,4.65) {ITCM $64$ KB};
  \draw (0,4.39) rectangle (\BW,4.91);
  \node[font=\small\itshape] at (0.5*\BW,4.65) {empty (by measurement)};
  \node[anchor=west, font=\footnotesize, text width=5.2cm] at (\BW+0.25,4.65)
    {kernels ran $11$--$15\%$ slower here; they live in cached \code{.text}};
  % ---- DTCM: full ----
  \node[anchor=east, font=\small\bfseries] at (-0.15,3.75) {DTCM $128$ KB};
  \draw[fill=green!14] (0,3.49) rectangle (\BW,4.01);
  \node[font=\small] at (0.5*\BW,3.75) {8 forward twiddle planes ($128$ KB exact)};
  \node[anchor=west, font=\footnotesize, text width=5.2cm] at (\BW+0.25,3.75)
    {copied + byte-verified at boot; $-6.4\%$ per forward NTT};
  % ---- Flash: 73% ----
  \node[anchor=east, font=\small\bfseries] at (-0.15,2.85) {Flash $511$ KB};
  \draw[fill=yellow!22] (0,2.59) rectangle (0.73*\BW,3.11);
  \draw (0,2.59) rectangle (\BW,3.11);
  \node[font=\small] at (0.5*\BW,2.85) {code $+$ rodata (incl.\ inverse twiddles)};
  \node[anchor=west, font=\footnotesize, text width=5.2cm] at (\BW+0.25,2.85)
    {${\approx}375$ KB; served through the I/D caches};
  % ---- AXISRAM: 32% peak live ----
  \node[anchor=east, font=\small\bfseries] at (-0.15,1.95) {AXISRAM $1536$ KB};
  \draw[fill=blue!13] (0,1.69) rectangle (0.318*\BW,2.21);
  \draw (0,1.69) rectangle (\BW,2.21);
  \node[font=\small] at (0.5*\BW,1.95) {per-prime polynomials (\code{.bss})};
  \node[anchor=west, font=\footnotesize, text width=5.2cm] at (\BW+0.25,1.95)
    {peak \emph{live} set $488.5$ KB (pk) / $408.5$ KB (sym)};
\end{tikzpicture}%
}
\vspace{-0.7cm}
\caption{Measured memory placement on the STM32N657 (reset-state TCM geometry; filled portion of each bar shows the fraction used). Annotations mark twiddle table placement, the ITCM experiment, and measured runtime peaks. Because the blocked NTT streams only one $256$-byte row tile per prime through the D-cache at a time, the design is cache-bound rather than memory-bound.}
\label{fig:memmap}
\end{figure}

\paragraph{Kernels to ITCM: measured and rejected.} Placing the NTT kernels' code in ITCM, the textbook embedded move, and originally our own plan too, made them $11$--$15\%$ \emph{slower} than simply executing the same bytes from cached AXISRAM-resident \code{.text} (forward \cyc{325314} vs.\ \cyc{284076} cycles; inverse \cyc{332008} vs.\ \cyc{298631}; identical kernel code, only its memory placement differs between the two measurements). The reason is that these kernels are fully unrolled, straight-line code, so the instruction cache streams them essentially perfectly on its own; on this silicon, the ITCM port turns out to simply be the slower fetch path. 
We therefore keep the kernels in ordinary cached \code{.text}, leaving ITCM empty, a result annotated as a measurement in Figure~\ref{fig:memmap}, not an oversight.

\subsection{The Helium $64\times64$ RNS NTT and Its Composition Audit}
\label{sec:ntt}

We factor the $4096$-point negacyclic NTT per prime as a blocked $64\times64$ (four-step) schedule: a transpose to make columns contiguous, $64$ calls to a $64$-point kernel, a transpose back, an inter-stage twiddle multiply, $64$ more kernel calls, and a final transpose. This factorisation is forced by size: each alternative fails on a measurable resource. Merging the whole transform in-register, as the $n=256$ PQC kernels do (Section~\ref{sec:related}), does not scale: a $16$~KB residue exceeds the $128$-byte register file by two orders of magnitude, forcing spills that forfeit exactly the memory traffic merging exists to avoid. A monolithic, fully-unrolled $4096$-point transform fails on code size instead: butterfly count grows as $(n/2)\log_2 n$, so such a kernel would cost ${\sim}128\times$ our $5$~KB $64$-point kernel, ${\sim}640$~KB per prime and direction against $511$~KB of total flash. A looped monolithic transform fits, but its later layers stride across the residue at large power-of-two distances, the access pattern a set-associative cache serves worst. Blocking avoids all three: every kernel call touches one contiguous $256$-byte row, the eight kernels together occupy $40$~KB and unroll completely, and strided access is confined to the three explicit transposes.

The $64$-point kernels, one forward and one inverse per prime, each generated with that prime's constants baked in, are fully unrolled, six-layer DIF/DIT butterfly networks in Helium assembly: each MVE operation processes four $32$-bit coefficients, the Montgomery product is the \code{vqrdmulh}/\code{vmul}/\code{vqrdmlah} triple, and every reduction is a straight-line, lane-wise mask. The kernels contain \emph{zero conditional branches}, verified on the disassembly, one \code{bx lr} each, so their timing is secret-independent by construction. In the final image, the forward transform spends $72\%$ of its time in the two kernel sweeps, $16\%$ in the vectorised twiddle pass, and $10\%$ in the three scalar transposes.

Proving the kernels themselves bit-exact was the easier half of this problem; the harder half is their \emph{composition}, since this is exactly where silent, self-consistent wrongness can hide. A dedicated audit of the full $4096$-point data flow found and fixed two such defects, a dropped inverse transpose, and inter-stage twiddle tables indexed in natural rather than bit-reversed row order. Both defects compute something other than the true negacyclic NTT, yet neither is visible from an ordinary encrypt-then-decrypt round-trip, since the same wrong bilinear map cancels itself out between encryption and decryption. Only a cross-implementation, byte-level test catches this class of bug, which is why the transform's composition, not just its individual kernels, is pinned by the KATs of Section~\ref{sec:correctness}. This same audit also established that the corrected forward transform is a \emph{pure permuted evaluation map}, the fact Section~\ref{sec:seeded} later turns into a zero-forward-NTT encrypt path, and confirmed that, once fixed, the forward and inverse kernels form an exact, unscaled transform pair with no compensating constants required anywhere.

\subsection{Lazy-Reduction Kernels}
\label{sec:lazy}

Each NTT butterfly adds and subtracts two values, and either result can grow slightly larger than the modulus $q_k$; left unchecked across many layers, this growth would eventually overflow the $32$-bit lane. \emph{Canonicalisation} is the fix: reducing a value back into $[0,q_k)$ after a butterfly, so it cannot grow further. The initial, correctness-first kernels (``v1'') canonicalise after \emph{every} layer, the safest possible choice, and the one under which the composition audit and the KATs were first established. It is also wasteful: canonicalising is not free, and most layers do not actually need it, values only overflow if they are allowed to grow unchecked for several layers in a row, so checking after every single layer does far more work than correctness requires, exactly the excess the PQC line already avoids~\cite{pqmx}. Our $27$-bit primes ($q_k<2^{27}$, values ${\approx}2^{26}$) leave enough headroom below the $32$-bit lane for a leaner schedule to be possible.

A \emph{lazy} schedule canonicalises only where growth could actually overflow, and skips it everywhere else. Getting this right safely, without silently overflowing somewhere we failed to check, needs a real proof, not a guess, so we built a range auditor that propagates exact per-stage magnitude bounds through the schedule, using the same instruction semantics as the composition audit, and the kernel \emph{generator refuses to emit} any schedule the auditor cannot certify safe. This analysis found that naive full laziness overflows at the fourth of the kernel's six butterfly layers, where the difference path doubles each layer, so the generator inserts one small, three-instruction correction there; the inverse transform, whose values grow more slowly (${\sim}1.5q$ per stage), runs fully lazy throughout, with its final multiply absorbing the remaining range. 
The certified worst case is $12.14q\approx2^{29.6}$, a $2\times$ margin under the $32$-bit lane (empirically, values never exceeded ${\sim}7q$ in testing); every output is still canonicalised to exactly $[0,q_k)$ before leaving the kernel. The kernels remain fully unrolled, straight-line, and branch-free, so the earlier constant-time argument still holds.

Removing safety checks is exactly the kind of change that could silently break correctness, so we required the leaner kernels to reproduce \emph{bit-identical} ciphertexts to v1: all five KATs serve as the acceptance gate, backed by $4000$ test vectors checking both kernel sets against each other on the emitted assembly. Measured on the same image before and after this change, the forward transform fell from \cyc{380777} to \cyc{285002} cycles per prime ($-25.2\%$), and the inverse from \cyc{475338} to \cyc{300421} ($-36.8\%$; the inverse dominates the seeded encrypt path), with emitted vector-instruction counts dropping $35\%$/$48\%$ respectively. Every downstream operation improved by exactly its NTT count times this per-prime saving, confirming the gain comes from the kernels alone.

\subsection{Seeded Ciphertexts}
\label{sec:seeded}
A BFV symmetric ciphertext is $(c_0,c_1)$ with $c_1=a$ uniformly random and public; like SEAL's own seeded mode, \lib{} transmits a $32$-byte PRNG seed in place of $c_1$, halving ciphertext size (Figure~\ref{fig:wire}), with the one added precondition that the error $e$ must stay unpredictable from that seed. We enforce this structurally, as discussed in Section~\ref{sec:threat}.

The on-device payoff is larger than the bandwidth saving alone. The composition audit (Section~\ref{sec:ntt}) established that the device's forward transform is a \emph{pure permuted evaluation map}: slot $i$ of the output holds the evaluation at $\psi^{2\,\mathrm{PERM}[i]+1}$, where $\psi$ is the primitive $2n$-th root of unity underlying the table $\Psi$ of Algorithm~\ref{alg:ntt} and $\mathrm{PERM}[i]=(\mathrm{br}_6(\lfloor i/64\rfloor)\ll6)\,|\,\mathrm{br}_6(i\bmod64)$, for every prime.
Uniform randomness is invariant under a fixed permutation, so the device samples $\hat a$ \emph{directly in the evaluation domain}, with the same bias-free rejection sampler, and the encrypt path needs \emph{zero forward NTTs}: for each prime, the device samples $\hat a$, performs one pointwise multiply by $\hat s$, and runs one inverse NTT. The receiver reconstructs $a=\mathrm{INTT}(\mathrm{unpermute}(\hat a))$ via \code{expand\_ct1}, proven bit-exact against the device's kernel composition (Section~\ref{sec:correctness}); public-key generation benefits identically ($8$ NTTs instead of $16$), and sampling $\hat a$ per prime also feeds the DMA prefetch of Section~\ref{sec:sampling}, so the two optimisations compose.

\begin{figure}[t]
\centering
\resizebox{0.75\linewidth}{!}{%
\begin{tikzpicture}[font=\small]
  % to-scale wire bars: 131072 -> 8.0cm ; 65572 -> 4.0016cm
  \node[anchor=west, font=\footnotesize\bfseries] at (0,1.62) {classic};
  \draw[fill=green!12] (0,1.05) rectangle (4.0,1.45);
  \node[znote] at (2.0,1.25) {$c_0$: 4 residues, $\num{65536}$ B};
  \draw[fill=orange!14] (4.0,1.05) rectangle (8.0,1.45);
  \node[znote] at (6.0,1.25) {$c_1$: 4 residues, $\num{65536}$ B};
  \node[anchor=west, font=\footnotesize] at (8.15,1.25) {$=\num{131072}$ B};
  \node[anchor=west, font=\footnotesize\bfseries] at (0,0.62) {seeded};
  \draw[fill=green!12] (0,0.05) rectangle (4.0,0.45);
  \node[znote] at (2.0,0.25) {\code{"BFV2"}\,$\|$\,$c_0$, $\num{65540}$ B};
  \draw[fill=black!25] (4.0,0.05) rectangle (4.10,0.45);
  \node[anchor=west, font=\footnotesize] at (4.25,0.25)
    {seed$_a$ ($32$ B) \quad $=\num{65572}$ B \ ($-50.0\%$)};
\end{tikzpicture}%
}
\vspace{-0.3cm}
\caption{Ciphertext transmission formats. The seeded format replaces the transmitted $c_1$ with a $32$-byte public seed that generates $\hat a$; a permutation identity (Section~\ref{sec:ntt}) lets the device sample $\hat a$ directly in the NTT domain, skipping the forward NTT on the encrypt path, while the receiver reconstructs $c_1$ deterministically from the seed.}
\label{fig:wire}
\end{figure}

\subsection{Further Optimisations}
\label{sec:tier2}

With the NTT, PRNG, and ciphertext transmission format settled, four scalar regions remained to optimise. We ported all four to the shared lane helpers of Section~\ref{sec:arith} together, each gated by the five KATs on the FVP and on silicon, plus a control build with the change flagged off. To confirm the improvements below are attributable to this change alone, we checked two things that should not have moved: the NTT, left untouched, moved by less than $0.2\%$, and key generation, also untouched, moved by only $+26$ cycles:
\begin{itemize}
  \item \textbf{Decrypt tail} ($-18.6\%$: \cyc{4676192} to \cyc{3805926}). The three post-NTT loops, pointwise $\odot\,\hat s$, the $v=c_0+c_1s$ canonicalisation, and the $\gamma$-correction scale-and-round (Section~\ref{sec:decrypt}), now run four-lane, with the $64$-bit accumulators split into $(\mathrm{lo},\mathrm{hi})$ lane pairs and carries applied by predication. 
  A dedicated test comparing the vector implementation against both the scalar C and the Python reference implementation confirms they agree exactly on $1.4$~million structured and random tuples. As a cross-check, a scalar-NTT decrypt variant carrying the same MVE tail fell by the same amount ($-\cyc{821044}$ vs.\ $-\cyc{870266}$ live), confirming the delta comes from the tail rather than any interaction with the NTT.
  \item \textbf{Batch encoder} ($-64.8\%$: \cyc{1503143} to \cyc{529381}). The mod-$t$ transform now uses per-layer twiddle tables in place of a serial root chain, \code{vld2q}/\code{vld4q} de-interleaved early stages, four-lane butterflies, and residue-0-only plaintext storage, since the encoded $m$ is written once rather than four times, with $\Delta m$ readers indexing residue $0$ directly. Building the tables costs \cyc{351804} cycles, once, at startup. What remains scalar is deliberately so: bit-reversal, a checked slot copy, and secure wipe.
  \item \textbf{Public-key path} (encrypt $-14.6\%$: \cyc{7636249} to \cyc{6521843}; key generation $-11.4\%$; seeded key generation $-20.4\%$). All eleven remaining scalar loops, the secret-$u$ lift, the dual pointwise $\hat u\odot\mathrm{pk}_{0,1}$, the ciphertext and public-key assembly windows in $[0,3q)$, and the dual Montgomery-form conversions, now reuse the same proven vector helpers as everything else in the library.
  \item \textbf{Gaussian CDT scan} (folded into the above; two sampler calls during public-key encryption, one during key generation, partially hidden behind DMA). This runs as a four-lane predicated scan, consuming keystream in the same order as before.
\end{itemize}

\subsection{Division-Free Decryption}
\label{sec:decrypt}

Decryption must compute $m=\lfloor t\smul X/Q\rceil\bmod t$ from $X=[c_0+c_1s]_Q$ (the noisy, unrounded decryption result before scaling), but $X$ exists only as four $27$-bit residues; the textbook route requires reconstructing a full $105$-bit value via CRT and then a $122$-bit division, both hostile to a $32$-bit core with no native support for arithmetic that wide. We adopt the BEHZ simple-scaling correction~\cite{behz16}, the same construction Microsoft SEAL implements with a $61$-bit auxiliary modulus, and \textbf{specialise the auxiliary modulus to $\gamma=2^{31}$}. This turns every ``$\bmod\,\gamma$'' into a $31$-bit \code{AND}, and the centred-lift test $u_\gamma\ge\gamma/2$ into simply reading bit~$30$; Algorithm~\ref{alg:decrypt} gives the resulting four-step, division-free procedure, computed per coefficient entirely in $32\times32{\to}64$ multiplies, adds, shifts, and masks.

The resulting datapath has \emph{no division and no secret-dependent branch}; all decrypt constants fit in $96$ precomputed bytes. Decryption is exact whenever the centred residue of $t\smul X\bmod Q$ stays below $Q(1/2-5/\gamma)$, that is, for any ciphertext retaining at least $2^{-28}$ of its noise budget (SEAL's $61$-bit $\gamma$ gives the analogous $2^{-58}$ caveat); both are comfortably irrelevant against our measured $72.3$-bit fresh budget. We checked the correctness of this specialisation against a big-integer $\lfloor t\smul X/Q\rceil\bmod t$ ground truth on $30{,}000$ cases, including worst-case-valid noise. The three decrypt loops described here also run four-lane in the final library (Section~\ref{sec:tier2}).

% =====================================================================
\section{Experimental Results}
\label{sec:eval}

\begin{table}[t]\centering
\caption{Evaluation platforms. Zeppelin runs on the STM32N657; the two SEAL-Embedded targets are included for context}
\vspace{-.3cm}
\label{tab:platform}
\resizebox{\linewidth}{!}{%
\begin{tabular}{@{}lllll@{}}
\toprule
Processor & Vector ISA & Clock & Cache & Memory \\
\midrule
\textbf{Cortex-M55 (ours)} & Helium/MVE & $800$\,MHz & I/D caches on &
ITCM $64$\,KB, DTCM $128$\,KB, AXISRAM $1.5$\,MB \\
Cortex-M4 (nRF52840) & none & $64$\,MHz & none & $256$\,KB \\
Cortex-A7 (MT3620) & NEON$^{\dagger}$ & $494$\,MHz & L1 $64/32$\,KB, L2 $256$\,KB & $256$\,KB$^{*}$ \\
\bottomrule
\end{tabular}%
}
\\[2pt]
\begin{minipage}{\linewidth}\scriptsize $^{\dagger}$NEON is present but
SEAL-Embedded's released NTT code is plain scalar C.\\$^{*}$The MT3620
reserves most RAM for its OS, leaving ${\sim}256$\,KB per application. TCM values are the measured reset-state geometry; FLEXMEM extension is unused.
\end{minipage}
\end{table}

\subsection{Experimental Setup, Parameters, and Security}
\label{sec:param}

We evaluate on the platform of Table~\ref{tab:platform}. The library is built with \code{arm-none-eabi-gcc} at \code{-O3} for the \code{armv8.1-m.main} architecture with \code{mve.fp} and hard float. Cycle counts are read from the Cortex-M55 \code{DWT\_CYCCNT} counter with instruction and data caches enabled and the PRNG served by the hardware CRYP peripheral through DMA (Section~\ref{sec:sampling}). Unless noted, every figure is the \textbf{warm median of $1000$ timed iterations} after one warm-up call, of the \emph{whole composed operation}; we deliberately do not sum per-stage instrumentation into our reported totals, since isolated stages cannot overlap the DMA keystream the way the composed operation does.

\paragraph{Parameters.}
$n=4096$, $t=65537$; four NTT-friendly primes $\{q_0,q_1,q_2,q_3\}=\{$67239937,
67280897, 67411969, 67452929$\}$, each $q_k\equiv1\pmod{8192}$ and
$2^{26}<q_k<2^{28}$, giving $\log_2|Q|=104.0$ bits (bit-length $105$).
Since $t\equiv1\pmod{2n}$, the plaintext ring gives $4096$ batch slots; benchmarks fill all $4096$. Neither encoding, a fixed $4096$-point transform, nor encryption depends on how many slots carry data, so the same figures cover SEAL-Embedded's CKKS workload of $n/2=2048$ values at equal ring degree (Section~\ref{sec:bfv}).
The theoretical fresh capacity is $\lfloor\log_2(Q/t)\rfloor\approx88$ bits; the \textbf{measured fresh invariant-noise budget is $72.3$ bits} (computed with SEAL's noise-budget formula, averaged across ciphertexts from both symmetric and public-key encryption; the worst single ciphertext observed was $72.2$ bits).
As a second, independent check, SEAL itself reports $73$ bits when reading an \emph{imported device ciphertext} directly (Section~\ref{sec:correctness}), two instruments agreeing on one quantity. A single ${\sim}30$-bit prime would afford only ${\sim}14$ bits, not enough for even one multiplication; four primes are required for HE standard 128-bit security.

\paragraph{Security.}
Our parameters sit \textbf{inside the HE standard's $128$-bit classical envelope}~\cite{hestd18}: at $n=4096$ with ternary secret and $\sigma=3.2$ the standard caps $\log_2 q$ at $109$ bits, and our $Q$ is $104.0$ bits, while SEAL's default $\{36,36,37\}$ at the same $n$ is itself $109$ bits, so our RLWE margin is no smaller than SEAL's at equal ring degree and distributions.
As corroboration, we ran the canonical lattice-estimator~\cite{latticeestimator} (SageMath~$10.9$) on our implemented instance ($n=4096$, $Q=104.0$ bits, uniform ternary secret, $\sigma=3.2$, $m=\infty$). Its full attack suite under the MATZOV cost model~\cite{matzov22} reports a cheapest attack of $\mathbf{2^{131.5}}$ operations (primal BDD at $\beta=348$; uSVP $2^{132.2}$, dual $2^{133.4}$, hybrid variants no cheaper given our dense secret), and its core-SVP quick estimate, $2^{101.9}$, matches the self-contained calibration of Bai--Galbraith ternary scaling~\cite{baigal14,adps16,aps15}, $2^{102.2}$, to within $0.3$ bits. The same suite applied to SEAL's own $109$-bit default at $n=4096$ reports $2^{125.1}$: our parameters measure $6.4$ bits above the deployed SEAL default under identical distributions, and clear $128$ bits under both the standard's gate-count metric and the newer MATZOV accounting (SEAL's sub-$128$ MATZOV reading reflects that model's stricter dual-sieve costs, not a defect in either parameter set).

\subsection{Per-Module Overhead}
\label{sec:eval-mod}

Table~\ref{tab:modules} reports the primitive modules. The NTT rows are the final lazy kernels (Section~\ref{sec:lazy}); the sampler rows carry the DMA-slab bimodality discussed in Section~\ref{sec:sampling}, so each sampler's minimum figure reflects an iteration that needed no slab refill. 
We finalize a single NTT configuration, precomputed Montgomery twiddles with forward planes resident in DTCM, rather than offering SEAL-Embedded's compute/load/load-fast spectrum of configuration options; our blocked scheduler already keeps the hot tile cache-resident no matter which of those options a developer might otherwise have needed to pick.

\begin{table}[t]\centering
\caption{Per-module cost on the Cortex-M55 ($n=4096$, warm median of $1000$; measured on the final library). $\times4$ rows cover all four primes.}
\vspace{-0.3cm}
\label{tab:modules}
\footnotesize
\begin{tabular}{@{}lrrl@{}}
\toprule
Module & Cycles & Time ($\mu$s) & Notes \\
\midrule
Forward NTT (1 prime)   & \cyc{284282} & $355$ & lazy kernels, DTCM twiddles \\
Inverse NTT (1 prime)   & \cyc{299198} & $374$ & \\
Forward NTT ($\times4$) & \cyc{1208785} & $1511$ & \\
Inverse NTT ($\times4$) & \cyc{1303916} & $1630$ & \\
Sample uniform $a$ ($\times4$) & \cyc{1712883} & $2141$ & rejection; public draw count \\
Sample ternary $s$      & \cyc{513276} & $642$ & CT; fastest observed \cyc{152087} (no refill needed) \\
Sample Gaussian $e$ ($\times4$) & \cyc{755298} & $944$ & CT 4-lane CDT; fastest observed \cyc{393489} \\
PRNG keystream          & \multicolumn{2}{l}{$28$ cyc/byte synchronous} & overlapped inside windowed ops \\
Batch encode (mod $t$)  & \cyc{529381} & $662$ & 4-lane; residue-0 store \\
Encode-table init       & \cyc{351804} & $440$ & one-time at startup \\
\bottomrule
\end{tabular}
\end{table}

\subsection{Isolating the Vectorisation Speedup (Same-Core Ablation)}
\label{sec:eval-ablation}

This section provides a controlled, same-core experiment that isolates the vectorisation speedup (the cross-library comparison in Section~\ref{sec:eval-comp}, by contrast, varies scheme, ISA, and core all at once, so it cannot attribute any speedup specifically to vectorisation). The experiment computes the identical $4096$-point negacyclic NTT with (a) the final Helium kernels and (b) a scalar implementation, on the \emph{same} Cortex-M55, with the same primes, the same Montgomery arithmetic, the same compiler, and \emph{in the same image}, so only the NTT implementation differs. The scalar baseline is a standard iterative Cooley--Tukey NTT, not a division-based strawman, and not hand-tuned: its translation unit is built with \code{-mcpu=cortex-m55+nomve} so the compiler cannot auto-vectorise it (verified directly: no MVE instructions appear in the disassembly), and, for fairness against kernels that load precomputed twiddles, it likewise loads precomputed per-stage twiddles in bit-reversed iteration order, so each butterfly costs exactly one Montgomery multiply (a running-product schedule instead would inflate the scalar cost by roughly $46\%$ and overstate the resulting speedup). The substitution preserves correctness as a whole unit: under fixed seeds, the scalar and Helium paths serialise to \emph{byte-identical} ciphertexts on silicon, confirmed by an identical FNV-1a digest.

Table~\ref{tab:ablation} is the paper's central speed evidence: on the same core, the vectorised engine is worth $2.55$--$2.79\times$ on the transform itself, and $1.84\times$/$2.09\times$ end-to-end on classic encrypt/decrypt. This difference is the combined contribution of Helium vectorisation and the blocking and lazy-reduction schedule built around it: against the original, pre-lazy-reduction kernels of Section~\ref{sec:lazy} the kernel-level ratio was $1.67$--$1.95\times$; the lazy schedule and the DTCM twiddle placement of Section~\ref{sec:memory} account for the rest. The four-lane MVE datapath sets a theoretical $4\times$ ceiling on the transform. Our measured $2.55$--$2.79\times$ falls short of that ceiling because the transform also spends time outside the vectorised butterflies, on three scalar transposes and the twiddle pass ($26\%$ of the forward transform, Section~\ref{sec:ntt}), and on the cache traffic of streaming $16$~KB residues, none of which the vector units accelerate. The scalar side of this comparison keeps the vectorised samplers, encoder, and decrypt tail unchanged, so these ratios isolate the NTT's contribution. Making those components scalar too, in a whole-program comparison, would show a larger overall speedup for Helium versus scalar, but it would mix several changes together instead of isolating one.

\begin{table}[t]\centering
\caption{Same-core ablation on the Cortex-M55 (same image, warm median of $1000$). The end-to-end (last-two) rows swap \emph{only} the NTT to scalar; everything else is unchanged.}
\vspace{-0.3cm}
\label{tab:ablation}
\footnotesize
\begin{tabular}{@{}lrrr@{}}
\toprule
Operation & Scalar (cyc) & Helium/MVE (cyc) & Speedup \\
\midrule
Forward NTT, 1 prime  & \cyc{794231} & \cyc{284282} & $2.79\times$ \\
Inverse NTT, 1 prime  & \cyc{798305} & \cyc{299198} & $2.67\times$ \\
Forward NTT, $\times4$ & \cyc{3340720} & \cyc{1208785} & $2.76\times$ \\
Inverse NTT, $\times4$ & \cyc{3318546} & \cyc{1303916} & $2.55\times$ \\
\midrule
Symmetric encryption & \cyc{9016463} & \cyc{4909613} & $1.84\times$ \\
Decryption & \cyc{7948514} & \cyc{3805926} & $2.09\times$ \\
\bottomrule
\end{tabular}
\end{table}

\subsection{End-to-End Cost}
\label{sec:eval-enc}

Table~\ref{tab:headline} reports the final cycle counts for each operation; Table~\ref{tab:campaign} shows the step-by-step history behind those numbers, tracking how each optimisation in Section~\ref{sec:impl} moved the library from its initial, unoptimised baseline to the final figures reported here. The seeded path is the deployment fast path (SEAL itself ships seeded symmetric ciphertexts); the public-key numbers are what cross-library comparisons should use, and we use them in Section~\ref{sec:eval-comp}.
Every row of Table~\ref{tab:campaign} was checked for reconciliation: each optimisation's measured gain had to be traceable to its actual cause (for instance, the decrypt-tail vectorisation of Section~\ref{sec:tier2} produces the same-sized improvement whether it is paired with the real Helium NTT or with a deliberately scalar NTT carrying the same vectorised tail, confirming the gain is the tail's alone), and any operation left untouched by a given change served as a null control, confirming it stayed put. The fastest individual runs confirm the same effect from a different angle: iterations that needed no slab refill at all completed in just \cyc{2903424} cycles ($3.63$\,ms), noticeably faster than the $4.10$\,ms median for seeded encryption. Both figures exclude batch encode.

\begin{table}[t]\centering
\caption{\lib{} core operations, measured on the STM32N657 ($n=4096$, four $27$-bit primes, $|Q|=104.0$ bits; warm median of $1000$; wall-clock at $800$\,MHz, including receiver-side \code{expand\_ct1}). Ciphertext sizes: classic \num{131072}\,B, seeded \num{65572}\,B. Each encode-and-encrypt row is the encryption median plus the batch-encode median, while the decrypt-and-decode row is the median of the composed operation.}
\vspace{-0.3cm}
\label{tab:headline}
\footnotesize
\begin{tabular}{@{}lrrr@{}}
\toprule
Operation & Cycles & Time (ms) & Energy ($\mu$J)$^{\dagger}$ \\
\midrule
Key generation & \cyc{2256053} & $2.82$ & $201.1$ \\
Batch encode (mod $t$) & \cyc{529381} & $0.66$ & $47.1$ \\
Batch decode (mod $t$) & \cyc{516199} & $0.65$ & $46.3$ \\
Symmetric encryption (seeded) & \cyc{3280971} & $4.10$ & $292.3$ \\
Symmetric encryption (classic) & \cyc{4909613} & $6.14$ & $437.7$ \\
Decryption ($\gamma$-correction, MVE tail; excludes batch decode) & \cyc{3805926} & $4.76$ & $339.4$ \\
Public-key encryption & \cyc{6521843} & $8.15$ & $581.1$ \\
Public-key generation & \cyc{8474082} & $10.59$ & $755.1$ \\
Public-key generation (seeded $\hat a$) & \cyc{4687768} & $5.86$ & $417.8$ \\
Receiver-side \code{expand\_ct1} (host/server) & \cyc{2605989} & --- & --- \\
\midrule
Symmetric encode-and-encrypt (seeded) & \cyc{3810352} & $4.76$ & $339.4$ \\
Symmetric encode-and-encrypt (classic) & \cyc{5438994} & $6.80$ & $484.8$ \\
Public-key encode-and-encrypt & \cyc{7051224} & $8.81$ & $628.2$ \\
Decrypt-and-decode & \cyc{4306711} & $5.38$ & $383.6$ \\
\bottomrule
\end{tabular}
\\[2pt]
{\footnotesize $^{\dagger}$Approximate core power, $71.3$\,mW at $0.81$\,V (STM32N657 datasheet~\cite{stm32n6ds}).}
\end{table}

\begin{table}[t]\centering
\caption{The sequence of optimisations, one per row, each gated by bit-exact KATs on the FVP and on silicon (Section~\ref{sec:correctness} defines the C1, C2, and W1 checks used in the KATs column). Cycles are warm medians of $1000$, measured at that step.}
\vspace{-0.3cm}
\label{tab:campaign}
\small
\resizebox{\linewidth}{!}{%
\begin{tabular}{@{}llrp{2.3cm}p{1.9cm}@{}}
\toprule
Step & Change & KATs & Classic symm. (Mcycles) & Seeded symm. (Mcycles) \\
\midrule
1 & Initial correctness-first baseline (Section~\ref{sec:impl}) & C1 &
  \barcellL{8.63}{2.00}{blue!35} & --- \\
2 & MVE ciphertext assembly and pointwise multiply; DMA preparation & C1 &
  \barcellL{7.15}{1.66}{blue!35} & --- \\
3 & CRYP$+$DMA keystream overlap (\S\ref{sec:sampling}) & C1 &
  \barcellL{6.15}{1.43}{blue!35} & --- \\
4 & Forward twiddles $\to$ DTCM (\S\ref{sec:memory}) & C1 &
  \barcellL{6.09}{1.41}{blue!35} & --- \\
5 & Seeded $\hat a$; compact ciphertext format (\S\ref{sec:seeded}) & C1-2 &
  \barcellL{6.02}{1.40}{blue!35} & \barcellR{4.02}{0.93}{ForestGreen!45} \\
6 & Range-proven lazy kernels (\S\ref{sec:lazy}) & C1-2,W1 &
  \barcellL{5.04}{1.17}{blue!35} & \barcellR{3.39}{0.79}{ForestGreen!45} \\
7 & Decrypt tail, encoder, public-key path, and CDT scan vectorised (\S\ref{sec:tier2}) & C1-2,W1 &
  \barcellL{4.91}{1.14}{blue!35} & \barcellR{3.28}{0.76}{ForestGreen!45} \\
\midrule
\multicolumn{3}{@{}l}{Net change vs.\ step 1} & $-43.1\%$ & $-62.0\%$ \\
\bottomrule
\end{tabular}%
}
\end{table}

\subsection{Comparison with SEAL-Embedded}
\label{sec:eval-comp}

SEAL-Embedded is the natural, and only prior, point of comparison for this specific benchmark: encode-and-encrypt of $2048$ packed values at $n=4096$ on MCU-class hardware, the one workload both libraries report (\lib{}'s benchmarks fill all $4096$ slots at identical cost, Section~\ref{sec:param}). Its published \emph{public-key} encode-and-encrypt costs $737$\,ms on a $64$\,MHz Cortex-M4 and $77$\,ms on a $494$\,MHz Cortex-A7~\cite{sealembedded}; Table~\ref{tab:compare} converts these to cycles (time $\times$ clock) as the clock-independent measure.

Cycle-for-cycle, our public-key encode-and-encrypt takes $5.4\times$ fewer cycles against the A7 and $6.7\times$ fewer against the M4, but this is an \textbf{uncontrolled} comparison: the figures bound a combined difference of scheme, ISA, implementation, and a newer core all at once; they do not isolate a single cause, which is why Section~\ref{sec:eval-ablation} exists. We headline cycles deliberately: wall-clock time adds a further clock-speed ratio of up to $12.5\times$ (against the M4) that says nothing about the libraries themselves. What these uncontrolled numbers do say, directionally rather than causally, is this: an embedded HE client built around BFV's integer transform in place of CKKS's double-precision inverse-FFT encoder, and computing that transform on a vector ISA, ends up $5.4$--$6.7\times$ cheaper in cycles than the only prior published embedded HE figures, before any clock-speed advantage is even considered.

\begin{table}[t]\centering
\caption{Public-key encode-and-encrypt at $n=4096$: $2048$ packed values for SEAL-Embedded, all $4096$ slots for \lib{} at identical cost (Section~\ref{sec:param}). \textbf{Not a controlled comparison}: scheme, ISA, and core all differ, and SEAL-Embedded's cycle counts are derived from its published, asymmetric-path times, making our public-key row the like-for-like pairing. Table~\ref{tab:ablation} isolates the vectorised NTT engine under controlled conditions.}
\vspace{-0.3cm}
\label{tab:compare}
\footnotesize
%\resizebox{\linewidth}{!}{%
\begin{tabular}{@{}lllrr@{}}
\toprule
Library / core & ISA & Scheme & Cycles & Time \\
\midrule
SEAL-Embedded, Cortex-M4 ($64$\,MHz) & scalar & CKKS & $47.2$\,M & $737$\,ms \\
SEAL-Embedded, Cortex-A7 ($494$\,MHz) & scalar$^{\dagger}$ & CKKS & $38.0$\,M & $77$\,ms \\
\lib{}, Cortex-M55 ($800$\,MHz)  $^{\ddagger}$ & Helium/MVE & BFV & $7.05$\,M & $8.81$\,ms \\
\quad Symmetric, classic format & Helium/MVE & BFV & $5.44$\,M & $6.80$\,ms \\
\quad Symmetric, seeded format & Helium/MVE & BFV & $3.81$\,M & $4.76$\,ms \\
\bottomrule
\end{tabular}%
%}
\\[2pt]
{\footnotesize $^{\dagger}$NEON present but not used for the NTT (the released code is plain scalar C).\\
$^{\ddagger}$\lib{} rows are encryption plus batch encode (\cyc{529381} cycles,
  Table~\ref{tab:headline}); the one-time encode-table initialisation (\cyc{351804} cycles) is excluded as amortised setup.}
\end{table}

\subsection{Memory Footprint}
\label{sec:eval-mem}

On this class of device, memory is not the constraint, it is a solved problem. The peak live set is $27$--$32\%$ of AXISRAM (Table~\ref{tab:mem}), the library allocates nothing dynamically, and the worst-case stack stays under $1.1$~KB; the real design problem is cache behaviour, not capacity. For context, desktop HE libraries allocate several megabytes for the same CKKS encode-and-encrypt at $n=4096$ (SEAL v3.6 uses ${\approx}4.98$\,MB by Valgrind's measure, $8.99$\,MB RSS, as reported by~\cite{sealembedded}), memory no microcontroller has to spare, which is exactly why purpose-built embedded libraries exist at all; \lib{}'s deliberate, non-dynamic memory management keeps it to a small fraction of that, on a different scheme and platform than the desktop comparison.

\lib{} uses more RAM than SEAL-Embedded's $64$--$137$\,KB \emph{by design}: we target a $1.5$\,MB vector MCU and spend full per-prime polynomials on keeping four-lane arithmetic fed, a latency-optimised design point rather than a minimum-RAM one. The measured profile shows this choice is not wasteful: the peak live set is $27$--$32\%$ of AXISRAM, and the pipeline is \emph{cache-bound, not memory-bound}, limited by how often small tiles must stream through the cache, not by whether the working set fits. Within one NTT, the twiddle tables are the only component that genuinely benefits from DTCM residency (Section~\ref{sec:memory}); the kernel code itself does not, since the instruction cache already streams these fully unrolled kernels efficiently on its own, which is why moving code into ITCM measured slower rather than faster. The footprint reported by the region rows is not a hard floor: that image was built for testing, and carries harness overhead a real deployment would not, six PRNG states, and the $256$\,KB of comparison tables the scalar ablation needs. A leaner, symmetric-only client is projected at roughly $470$\,KB of statically allocated AXISRAM \code{.bss}, larger than Table~\ref{tab:mem}'s $408.5$\,KB symmetric live-set figure because the two measure different things: the projection totals every buffer the client links, since all state is allocated statically for the program's lifetime, while the live-set row counts only buffers concurrently in use at symmetric encrypt's peak. The ${\sim}60$\,KB gap is lifetime-disjoint state, chiefly the serialization buffer and the encoder's tables, which a tighter client could reuse for other purposes since it is never needed at that peak.

\begin{table}[t]\centering
\caption{Measured memory on the STM32N657. Region rows are the test image as linked and booted, larger than a real deployment needs; live-set rows are per-stage peaks from the instrumented profile, the figures a client should actually provision.}
\vspace{-0.3cm}
\label{tab:mem}
\footnotesize
%\resizebox{\linewidth}{!}{%
\begin{tabular}{@{}llrr@{}}
\toprule
 & Contents & Used & Capacity \\
\midrule
ITCM & \emph{empty by measurement} (\S\ref{sec:memory}) & $0$ & $64$\,KB \\
DTCM & 8 forward twiddle planes (byte-verified) & $128$\,KB & $128$\,KB \\
Flash & code $+$ rodata (incl.\ inverse twiddles) & ${\approx}375$\,KB & $511$\,KB \\
AXISRAM & benchmark image \code{.bss} (incl.\ harness duplicates) & \cyc{1310784}\,B & $1536$\,KB \\
\midrule
live set & public-key encrypt (whole-run peak) & $488.5$\,KB & \\
live set & symmetric-only client & $408.5$\,KB & \\
live set & decrypt-only client & $272$\,KB & \\
live set & serialize/deserialize transient (ciphertext bytes) & $256$\,KB & \\
stack / heap & worst-case HWM (silicon) / library allocations & \cyc{1056}\,B / $0$ & $2$\,KB reserve \\
\bottomrule
\end{tabular}%
%}
\end{table}

\subsection{Correctness and SEAL Interoperability}
\label{sec:correctness}

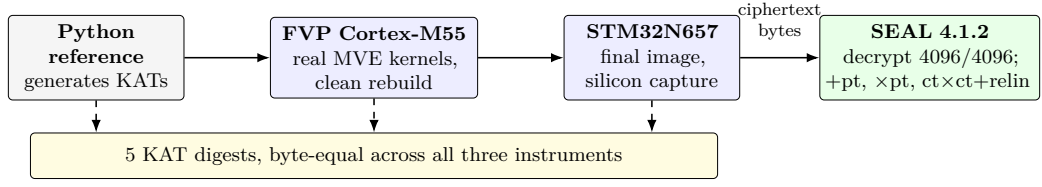
\begin{figure}[t]
\centering
\resizebox{\linewidth}{!}{%
\begin{tikzpicture}[font=\footnotesize]
  \node[zbox, text width=2.5cm] (py) at (0,0)
    {\textbf{Python reference}\\generates KATs};
  \node[zop, text width=3cm] (fvp) at (4.4,0)
    {\textbf{FVP Cortex-M55}\\real MVE kernels,\\clean rebuild};
  \node[zop, text width=2.5cm] (board) at (8.8,0)
    {\textbf{STM32N657}\\final image,\\silicon capture};
  \node[zpub, text width=3.2cm] (seal) at (13.2,0)
    {\textbf{SEAL 4.1.2}\\decrypt $4096/4096$;\\
     $+$pt, $\times$pt, ct$\times$ct$+$relin};
  \draw[zarr] (py) -- (fvp);
  \draw[zarr] (fvp) -- (board);
  \draw[zarr] (board) -- node[above, yshift=4.5pt, align=center, font=\footnotesize]{ciphertext\\bytes} (seal);
  \node[zmem, text width=10.6cm, minimum height=0.72cm] (kat) at (4.4,-1.55)
    {5 KAT digests, byte-equal across all three instruments};
  \draw[zret] (py.south) -- (kat.north -| py.south);
  \draw[zret] (fvp.south) -- (kat.north -| fvp.south);
  \draw[zret] (board.south) -- (kat.north -| board.south);
\end{tikzpicture}%
}
\vspace{-0.5cm}
\caption{The verification chain. The same five FNV-1a hashes are required of the reference implementation, the instruction-accurate FVP running the real kernels, and the silicon; the same serialized bytes then import into unmodified SEAL for decryption and evaluation.}
\label{fig:validation}
\end{figure}

\paragraph{Four instruments, five KATs.}
Correctness rests on four independent instruments: (i) a bit-exact Python reference implementation of the C pipeline, serving as the specification; (ii) an instruction-accurate model of the MVE kernel semantics; (iii) execution of the \emph{real} kernels on the ARM FVP Corstone SSE-300 Cortex-M55 model~\cite{fvp}; and (iv) big-integer ground truth for the $\gamma$-decryption. These instruments are checked against each other via FNV-1a hashes of five serialized artifacts: (1) the plaintext, (2) the classic symmetric ciphertext, and (3) the public-key ciphertext, together \textbf{C1}; and (4) the seeded symmetric ciphertext and (5) the seeded public-key chain, together \textbf{C2}, since both depend on the seeded-format optimisation of Section~\ref{sec:seeded}. A separate check, \textbf{W1}, confirms byte-identical output between the scalar and Helium/MVE implementations once both exist to compare (Section~\ref{sec:eval-ablation}); Table~\ref{tab:campaign} uses all three labels. The C2 hashes were generated from the Python reference before any C code for the seeded format existed, and only later required to match on the FVP and the board. A broken NTT composition still \emph{passes} an on-device encrypt-decrypt round-trip (Section~\ref{sec:ntt}); only this cross-implementation, byte-exact check catches it. All five KATs pass on the real device, along with the seeded round-trip and the scalar-versus-Helium ciphertext match from Section~\ref{sec:eval-ablation}, in a single test run on the final image, and the FVP independently reproduces the same five results from a clean rebuild. Figure~\ref{fig:validation} shows this verification chain.

\paragraph{SEAL interoperability, demonstrated.}
The cloud adapter reformats the device's ciphertext bytes into what unmodified Microsoft SEAL $4.1.2$ expects, range-checking and copying each value into SEAL's residue layout; the adapter itself never touches the secret key, which SEAL converts separately, using its own key-level NTT, before decrypting under it. We confirm three things: First, SEAL's \textsf{Decryptor} recovers the device's plaintext exactly, all $4096$ coefficients, for both the classic and seeded formats; the adapter's reconstruction of the mask $\hat a$ from its seed matches our reference implementation's before decryption even begins. Second, SEAL correctly evaluates on device ciphertexts using plain (unencrypted) operands: adding the plaintext $m$ to an encrypted ciphertext of $m$ gives $2m$ once decrypted, and multiplying by the plaintext $3$ gives $3m$, with zero mismatches across all $4096$ coefficients. Third, and most demanding, \textbf{a full multiplication succeeds}: using the relinearization keys described in Section~\ref{sec:threat}, the evaluator squares a device ciphertext and relinearizes it; the result decrypts to exactly $m^2$, with the noise-budget reading dropping from $73$ bits fresh to $46$ bits after the multiplication, comfortable headroom, not just a bare pass. Ciphertexts from \lib{} are, in every practical sense, ordinary SEAL ciphertexts.

\subsection{Timing Side-Channel Discussion}
\label{sec:ct}

We claim \emph{timing} resistance, argued at the C-source and ISA level and verified on the emitted assembly; power/EM and fault attacks are out of scope. (i) The ternary and Gaussian samplers consume a fixed amount of keystream with no branch on the values they produce; the uniform sampler rejects only on the public $a$/$\hat a$. (ii) Every secret-dependent reduction is a branch-free mask. This claim rests on a found-and-fixed leak, not just review: the RNS Montgomery helper's \code{if (r $\ge$ q) r -= q}, on paths touching $\hat a\odot\hat s$ and $c_1\smul s$, up to $16384$ invocations per operation, compiled to a conditional branch (\code{bcc}); the branch-free mask form we now use compiles straight-line, is bit-identical, and left every KAT unchanged. (iii) All eight of the final lazy kernels are fully unrolled with zero conditional branches, one \code{bx lr} each, with no predication used as control flow. (iv) The vectorised code of Section~\ref{sec:tier2} replaces branches with lane predication, both to select coefficients and to apply carries in the split accumulators; some of these predicates are computed from secret values, but a predicated MVE instruction takes the same time whether its predicate is true or false, and a predicate never governs branching or addressing, so timing stays independent of secret data~\cite{m55trm}. (v) The DMA keystream machinery branches only on slab geometry and call order, never on secret values, and the PRNG itself is the hardware AES engine, which uses no table lookups. The only branches remaining in secret-handling code are loop back-edges and slab-refill tests on public quantities. Decryption and decoding are division-free end to end.

% =====================================================================
\section{Conclusion}
\label{sec:concl}

\lib{} is the first HE library to exploit an embedded vector ISA and the first BFV client library, encoding, key generation, encryption, and decryption all on the device, with evaluation server-side, on MCU-class hardware. A blocked Helium RNS NTT over range-proven lazy kernels, a seeded ciphertext format built on an audited permutation identity, and a division-free $\gamma=2^{31}$ decryption deliver symmetric encode-and-encrypt in $4.76$\,ms and decrypt-and-decode in $5.38$\,ms at $n=4096$, with a $104$-bit modulus and a measured $72.3$-bit budget, a $62\%$ reduction in symmetric encryption cycles from where the correctness-first system started, every step KAT-gated on silicon. A controlled same-core ablation puts the vectorised engine at ${\approx}2.7\times$ per transform over an equivalent scalar NTT, and unmodified Microsoft SEAL decrypts and evaluates the device's ciphertexts through a depth-1 relinearized multiplication. The library is constant-time against timing channels, verified on emitted assembly. Power/EM evaluation and a SLOTHY scheduling pass over the lazy kernels are future work.

\section*{Acknowledgements}
LLMs were used for grammar and readability checks during manuscript preparation.

% =====================================================================
\appendix
\section{Algorithms}
\label{app:algos}
\begin{algorithm}[h]
\caption{In-place iterative negacyclic NTT (forward), per prime $q$. $\Psi$: precomputed bit-reversed $2n$-th roots in Montgomery form. This is the simple, textbook version; the final engine instead computes the blocked $64\times64$ schedule of Section~\ref{sec:ntt} with the lazy in-kernel reductions of Section~\ref{sec:lazy}, but produces exactly the same output, bit for bit.}
\label{alg:ntt}
\begin{algorithmic}[1]
\Function{NTT}{$a,\Psi,n,q$}
  \For{$(m=1;\ m<n;\ m=2m)$}
    \State $t \gets n/(2m)$
    \For{$(i=0;\ i<m;\ i{+}{+})$}
      \State $S \gets \Psi[m+i]$
      \For{$(j=2it;\ j<(2i{+}1)t;\ j{+}{+})$}
        \State $L \gets a[j]$;\quad $R \gets \textsc{MontMul}(a[j+t],S,q)$
        \State $a[j] \gets \textsc{Canon}(L+R,q)$
        \State $a[j+t] \gets \textsc{Canon}(L-R,q)$ \Comment{branch-free mask}
      \EndFor
    \EndFor
  \EndFor
  \State \Return $a$
\EndFunction
\end{algorithmic}
\end{algorithm}

\begin{algorithm}[h]
\caption{Division-free BEHZ scale-and-round decryption with
$\gamma=2^{31}$, per coefficient. Inputs: residues $x_k=[c_0+c_1 s]_{q_k}$.
Precomputed constants (Section~\ref{sec:decrypt}):
$\theta_k=[t\gamma\,(Q/q_k)^{-1}]_{q_k}$, stored in Montgomery form;
the base-conversion weights $C_{t,k}=[-Q^{-1}(Q/q_k)]_t$ and
$C_{\gamma,k}=[-Q^{-1}(Q/q_k)]_\gamma$, with BEHZ's $-Q^{-1}$ factor
folded in; and $\gamma^{-1}$, the inverse of $\gamma$ modulo $t$.}
\label{alg:decrypt}
\begin{algorithmic}[1]
\Function{Decrypt1}{$x_0,\dots,x_{L-1}$}
  \For{$k=0$ to $L-1$}
    \State $h_k \gets \textsc{MontMul}(x_k,\theta_k,q_k)$
  \EndFor
  \State $u_t \gets [\,\textstyle\sum_k h_k C_{t,k}\,]_t$
  \State $u_\gamma \gets (\textstyle\sum_k h_k C_{\gamma,k}) \mathbin{\&} (2^{31}-1)$
  \State $E \gets u_\gamma - \gamma\cdot\mathrm{bit}_{30}(u_\gamma)$
        \Comment{branch-free centred lift}
  \State \Return $[(u_t - [E]_t)\cdot\gamma^{-1}]_t$
\EndFunction
\end{algorithmic}
\end{algorithm}

% =====================================================================
\bibliographystyle{alpha}
\bibliography{zeppelin}

\end{document}